\documentclass[twocolumn,showpacs,showkeys,amsmath,amssymb,groupedaddress,aps,mcite,mciteplus,floatfix]{revtex4-1}

\usepackage{graphicx}
\usepackage{color}
\usepackage{float}

\begin{document}

\title{Disorder-induced cubic phase in Fe$_2$-based Heusler alloys.}

\author{Janos~Kiss}
\author{Stanislav~Chadov}
\author{Gerhard~H.~Fecher}
\email{gerhard.fecher@cpfs.mpg.de}
\author{Claudia~Felser}
\affiliation{Max-Planck-Institut f\"ur Chemische Physik fester Stoffe,\\ N\"othnitzer Strasse~40, 01187~Dresden, Germany.}


\begin{abstract}
Based on first-principles electronic structure calculations,
we analyze the chemical and magnetic mechanisms stabilizing the cubic phase in Fe$_2$-based
Heusler materials, which were previously predicted to be tetragonal when being
chemically ordered. In agreement with recent experimental data, 
we found that these compounds crystallize within the so-called ``inverted'' cubic Heusler
structure  perturbed by a certain portion of the intrinsic chemical
disorder. Understanding these mechanisms is a necessary step to
guide towards the successful future synthesis of the stable Fe$_2$-based
tetragonal phases, which are very promising candidates for the fabrication of rare-earth-free permanent magnets.
\end{abstract}

\pacs{}

\keywords{rare-earth-free, hard magnets, Heusler alloys, chemical disorder} 

\maketitle

One of the oldest problems within the field of materials science is
the search for the inexpensive hard magnets, i.\,e., for materials
retaining their magnetisation after being once magnetized.
Their role in the daily life can be hardly overestimated: hard magnets are widely
used in automotive applications, telecommunications, data processing,
consumer electronics, instrumentation, aerospace and bio-surgical
applications.  In particular, they play a unique role in
renewable energy technologies based on electric generators (e.\,g.,
rotors in wind-turbines, small hydroelectric systems etc.). 
However, materials exhibiting outstanding hard-magnetic properties together
with high magnetization and high Curie
temperature are rather expensive as being based on combinations
involving rare-earth elements (e.\,g., Sm-Co, Nd-Fe-B)~\cite{Kir96,Coe10}. Thus, the
development of new inexpensive compounds with hard-magnetic properties 
(i.\,e., rare-earth-free hard magnets) which can be industrially mass produced 
is important and highly relevant (for the review see e.\,g., Ref.~\cite{Coe10}). 
The recent explosion of attention for the tetragonally-distorted magnetic
Heusler systems originates at a large extent from this prospective as well~\cite{WCG+12}.

Indeed, apart of being promising candidates for tunneling
magneto-resistance and spin-torque-transfer applications~\cite{KMW+11,*JFB+11,WCG+12,*MKM+12,*CKF12,*OKF+12}, 
this family may also provide materials combining the tetragonal distortion  with a large magnetic
moment and high Curie temperature, suitable as hard magnets. The group
of Fe$_2$YZ-based Heusler compounds (with Y and Z being the transition
and the main-group element, respectively), theoretically predicted to be
tetragonal with a large  magnetization (4--5~$\mu_{\rm  B}$/f.u.,
f.u.=formula unit) would be one of such promising materials sources~\cite{GD08,*GD09}.
In contrary, the subsequent synthesis, XRD and M\"ossbauer characterization
have shown that all these compounds crystallize in the cubic phase~\cite{GKK+12}.
To understand which ingredients can lead to their tetragonal distortion
obviously implies an important preliminary step -- a detailed
understanding  of the mechanisms stabilizing their cubic phase. This is the
main point of the present study. 

\begin{figure}
\centering
\includegraphics[clip,width=0.7\linewidth]{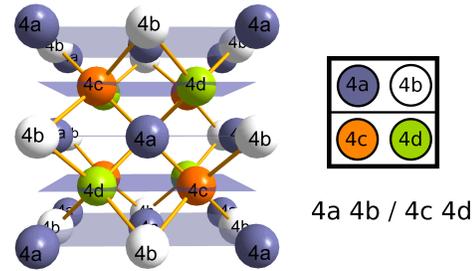}
\caption{(color online) The cubic unit cell of the 
point-symmetry group No.~216 together with its schematic graphical diagram and
  the corresponding written notation. The high-symmetric Wyckoff
  positions $4a$, $4b$, $4c$ and $4d$ are distinguished by different colors.
\label{FIG:notations}}
\end{figure}

Before proceeding to the results, first we would like to introduce the 
notations extensively used throughout in the text (see Fig.~\ref{FIG:notations}). 
In the most general case any cubic Heusler system corresponds at least
to the point-symmetry group No.~216. In order to distinguish between different chemical 
configurations, we will use the special written notation according to the
occupations of the four high-symmetric Wyckoff positions: first we will write down the
occupants of the $4a$ and $4b$ sites followed by the slash sign, then - of
the $4c$ and $4d$, i.\,e.,  ``$4a\,4b$/$4c\,4d$''. Thus, e.\,g. the
so-called ``regular'' and ``inverted'' variants (terms introduced in Ref.~\cite{GD08,*GD09}) of Fe$_2$CuGa can be
written as CuGa/FeFe and FeGa/CuFe, respectively. In case if certain
Wyckoff position is occupied by several atomic sorts randomly,
e.\,g. by A and B with probabilities $x$ and $1-x$, it is noted using the square
brackets: [A$_x$B$_{1-x}$]. In case of the tetragonal distortion (${c\neq
a}$) the symmetry reduces at least to the point group No.~119. The
sequence for the written notation in this case does not change, it 
implies only the usage of the different Wyckoff positions: $2a\,2b$/$2c\,2d$.

To clarify the discrepancy between the theoretical predictions from
Ref.~\cite{GD08,*GD09} and the experimental data~\cite{GKK+12}, in the following we will
study the relative stability of the cubic and tetragonal phases of
the Fe$_2$-based Heusler systems by optimizing both structural, magnetic and
chemical degrees of freedom based on {\it ab-initio} density-functional calculations.
As a suitable numerical tool which accounts for these
factors simultaneously, we use the fully-relativistic Green's function formalism
implemented within the SPR-KKR (Spin-Polarized fully Relativistic Korringa-Kohn-Rostoker) method~\cite{EKM11}. The random
occupation is described in terms of the CPA (Coherent Potential
Approximation)~\cite{Kor58,*Sov67,*Gyo72,*But85}. Despite its
mean-field nature (the effective averaging of the short-range order
effects) CPA remains the most practical technique which includes the 
essential features of randomness.
In order to ensure that the CPA result is not an artifact of the single-site approximation,
we performed additional supercell calculations. 
It was also found, that the usage of the full potential (i.\,e., the non-spherical potential) is much
more essential for the adequate description than a particular choice of the exchange-correlation
potential. For this reason, the presented calculations corresponds to
the fully-relativistic and full-potential results employing the local
density approximation for the exchange-correlation functional~\cite{VWN80}.
The calculations for different $c/a$ ratios
are performed for the fixed volume, which was taken from the available 
experimental data~\cite{GKK+12}.

In order to explain the mechanism which keeps the Fe$_2$-based
systems cubic, throughout the discussion we will focus on the Fe$_2$CuGa
system, because as we found, all basic conclusions valid for this
system can be transferred without restrictions onto other compounds
in this series (i.\,e., Fe$_2$CuAl, Fe$_2$NiGa, Fe$_2$NiGe and Fe$_2$CoGe)
synthesized experimentally~\cite{GKK+12}.

\begin{figure}
\centering
\includegraphics[clip,width=0.48\linewidth]{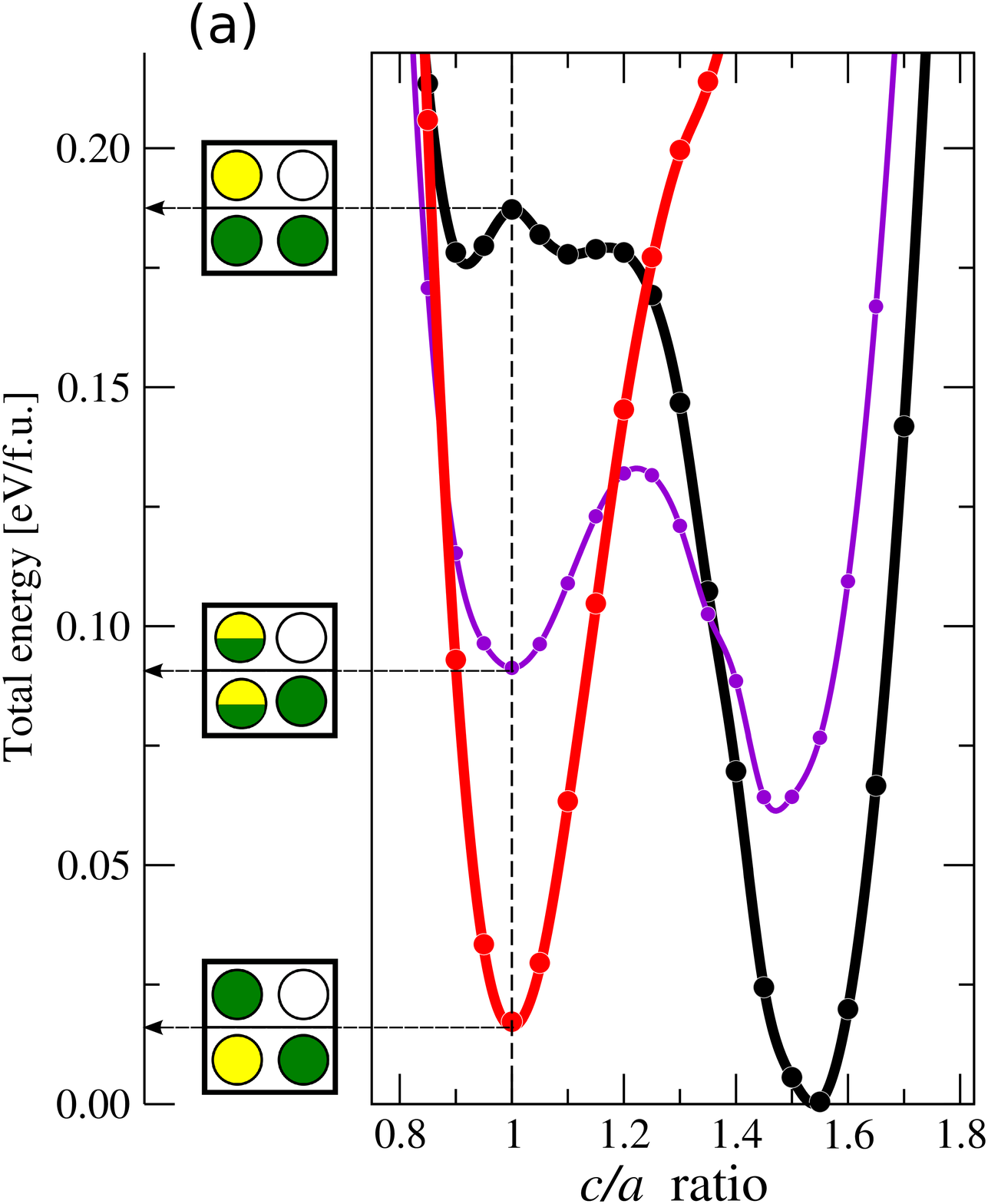}~~\includegraphics[clip,width=0.48\linewidth]{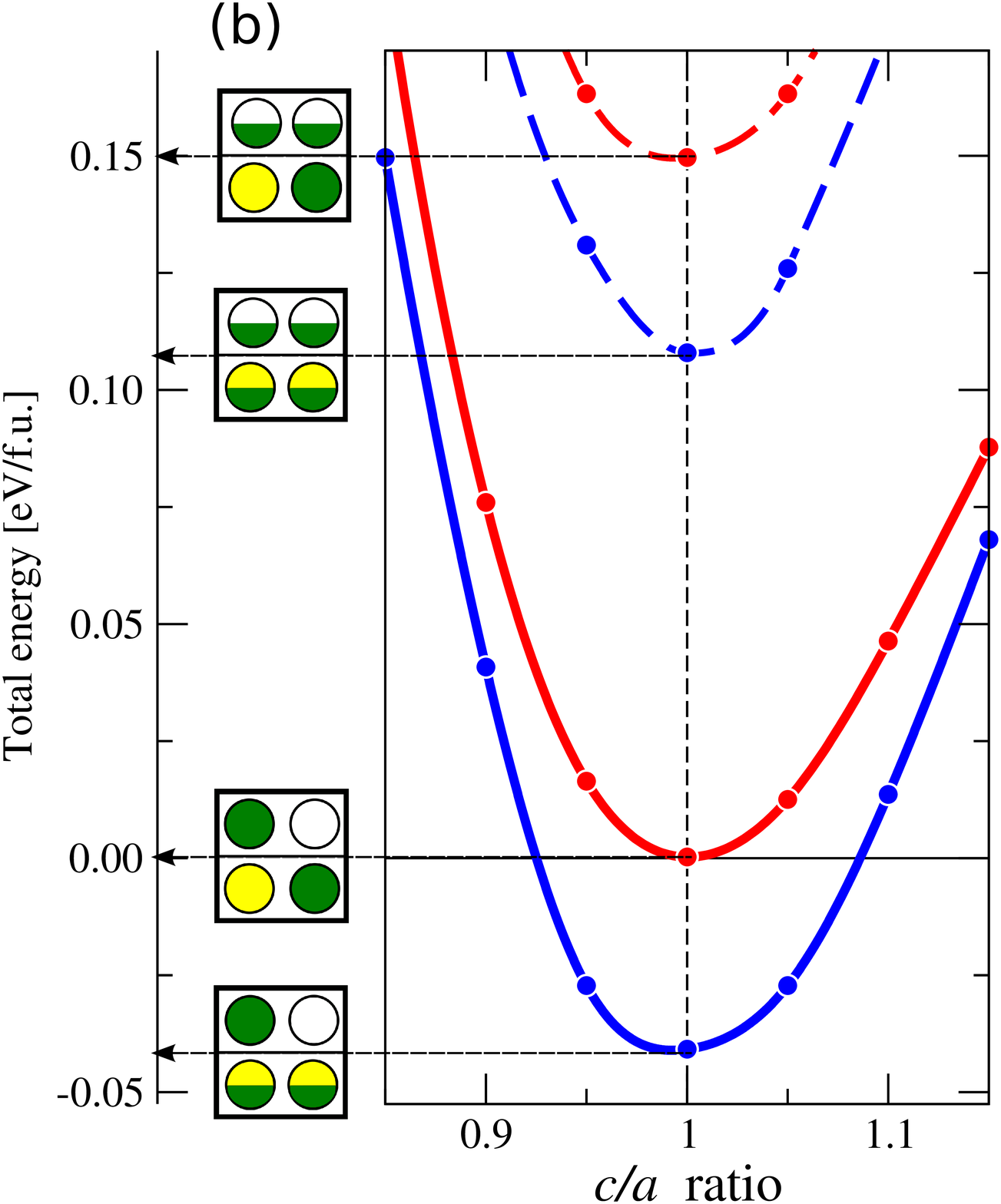}
\caption{(color online) Total energy of Fe$_2$CuGa Heusler alloy calculated as a function of $c/a$ ratio for
  various distributions  of Fe and Cu (indicated by the box-like diagram; green,
  yellow and white colored areas correspond to the Fe, Cu and Ga occupations, respectively). 
  (a)~Thick black and red curves correspond to CuGa/FeFe (``regular'')
  and  FeGa/CuFe  (``inverted'') configurations, respectively. Thinner
   violet line shows the intermediate  [Cu$_{0.5}$Fe$_{0.5}$]Ga/[Cu$_{0.5}$Fe$_{0.5}$]Fe case  between ``regular'' and ``inverted''. The absolute energy minimum of the ``regular'' (CuGa/FeFe) tetragonal phase
is taken as a reference. (b)~Solid red curve represents the same ``inverted''
 configuration as on the left panel, whereas  the solid 
blue line shows the most stable configuration:
FeGa/[Cu$_{0.5}$Fe$_{0.5}$][Cu$_{0.5}$Fe$_{0.5}$], 
obtained from the ordered ``inverted'' by mixing Fe and Cu randomly in-plane.
The dashed red and blue lines correspond to the distributions derived
from the previous two configurations via additional in-plane random
spread of Ga and Fe: [Fe$_{0.5}$Ga$_{0.5}$][Fe$_{0.5}$Ga$_{0.5}$]/CuFe
(red dashed) and [Fe$_{0.5}$Ga$_{0.5}$][Fe$_{0.5}$Ga$_{0.5}$]/[Cu$_{0.5}$Fe$_{0.5}$][Cu$_{0.5}$Fe$_{0.5}$]
(blue dashed). The absolute energy minimum of the ``inverted'' (FeGa/CuFe) phase
is taken as a reference.
\label{FIG:energies}}
\end{figure}

The main outcome of the present study is summarized in
Fig.~\ref{FIG:energies}, which represents the dependency of the total energy 
on $c/a$ ratio for various alloy configurations. 
The ordered ``regular'' Heusler structure in cubic phase turns out to be
unstable (indicated by the corresponding energy curve maximum at
${c/a=1}$, black line in Fig.~\ref{FIG:energies}\,(a)), whereas at about ${c/a=1.54}$ the system falls
into the relatively deep energy minimum. For the fixed chemical order (i.\,e., CuGa/FeFe) the tetragonal distortion is the only
mechanism which can relax the instability of the cubic phase, since the
magnetic degrees of freedom are already in use (for more detailed
description of tetragonal distortion mechanisms, see e.\,g.~\cite{GD08,*GD09,WCG+12}). 
As it follows, the gradual transition towards the ``inverted''
Heusler structure, realized by random chemical Cu-Fe inter-layer exchange
starts gradually to develop an energy minimum for the cubic phase.  
Although the configurations with intermediate Cu-Fe site occupations
(e.\,g., [Cu$_{0.5}$Fe$_{0.5}$]Ga/[Cu$_{0.5}$Fe$_{0.5}$]Fe) exhibit an
energy minima for both tetragonal and cubic phases, the limiting ordered system
(${x=1}$, i.\,e. the fully ``inverted'' FeGa/CuFe) is stable only within the
cubic phase (Fig.~\ref{FIG:energies}~(a) or (b), red curve).
Despite the large energy difference (about $-170$~meV/f.u.\@) 
between the cubic ``regular'' and ``inverted'' phases, the deepest absolute
energy minimum is found for the tetragonally-distorted ``regular'' configuration 
(see Fig.~\ref{FIG:energies}\,(a)), which is by about 20~meV/f.u.\@ more
stable compared to the ``inverted'' cubic configuration. 
Thus, for the ordered systems our results agrees with the former calculations~\cite{GD08,*GD09}. 
This means, that the mechanisms stabilizing the cubic phase involve degrees of freedom which 
where neglected so far, e.\,g. the chemical disorder~\cite{GKK+12}.

It is important to note, that the huge energy decrease (about
$-170$~meV/f.u.) gained by going from the ``regular'' cubic to the ``inverted'' cubic
configuration (the largest energy scale in the diagram on Fig.~\ref{FIG:energies}) 
is most likely of the magnetic origin. 
The latter is due to the optimization of the magnetic exchange
coupling within the Fe sublattice, since the nearest magnetic neighbors
(i.\,e., the Fe atoms from the adjacent layers within ``inverted'' cubic FeGa/CuFe) 
are sitting closer to one another compared to the ``regular'' cubic CuGa/FeFe setup, in
which they are in-plane. For this reason, by  searching for the more stable configurations,
we start from the ``inverse'' cubic system and perturb it by in-plane chemical disorder 
(i.\,e., by conserving the total amount of Cu and Ga within adjacent layers). 
Hence, there are two important in-plane disorder scenarios: random in-plane
mixtures of Fe-Ga, and that of Fe-Cu.

We found, that the random in-plane spread of Ga and Fe 
(case [Fe$_{0.5}$Ga$_{0.5}$][Fe$_{0.5}$Ga$_{0.5}$]/CuFe, red dashed line in Fig.~\ref{FIG:energies}\,(b)) 
leads to the increase of the total energy (compared to the ``inverted'' configuration,
FeGa/CuFe) by about 150 meV.
In contrast, the random in-plane spread of Cu and Fe (case FeGa/[Cu$_{0.5}$Fe$_{0.5}$][Cu$_{0.5}$Fe$_{0.5}$],
solid blue line in Fig.~\ref{FIG:energies}\,(b)) leads to an energy gain of about $-40$~meV 
(again, compared to the ``inverted'' case, FeGa/CuFe). 
The key observation is, that this $-40$~meV energy gain is enough to stabilize the cubic structure 
(in FeGa/[Cu$_{0.5}$Fe$_{0.5}$][Cu$_{0.5}$Fe$_{0.5}$ configuration),
which finally becomes more stable than the tetragonal ``regular'' ordered
CuGa/FeFe by about ${40-20=20}$~meV/f.u., in agreement with experiment.

Our results show, that these two effects (Fe-Ga and Fe-Cu in-plane random mixtures) 
are rather independent on one another: i.\,e.\@ disregarding the particular arrangement
of atoms within the adjacent layer, the energy changed by 150, 40 or ${150\pm40}$~meV while 
going from one distribution to another within all four cases. 
The large increase in energy by 150~meV in the first case is mainly due to the 
distinct nature of Fe and Ga.
So, within the fixed square lattice it is unfavorable to form separate clusters of Fe and Ga, 
since each atomic sort would prefer to create its own lattice within a cluster
which will be rather different from another. For this reason, any perturbation of the
perfect chemical order in Fe-Ga layers will increase the total energy. 
This issue, however, is not critical for the second case: 
the separation of Fe and Cu within the given lattice does not cost so much
energy, since both atom types are much more similar.  
To ensure that the $-40$~meV energy gain in this case
is not just an artifact of the single-site nature of the CPA, we have
performed supercell calculations by systematically increasing the
number of Fe-Cu in-plane swaps, mimiquing Fe-Cu disorder. 
This has shown that by increasing the degree of Fe-Cu separation the total energy 
is indeed reduced by around $-40$~meV.

\begin{figure} 
\centering
\includegraphics[clip,width=0.7\linewidth]{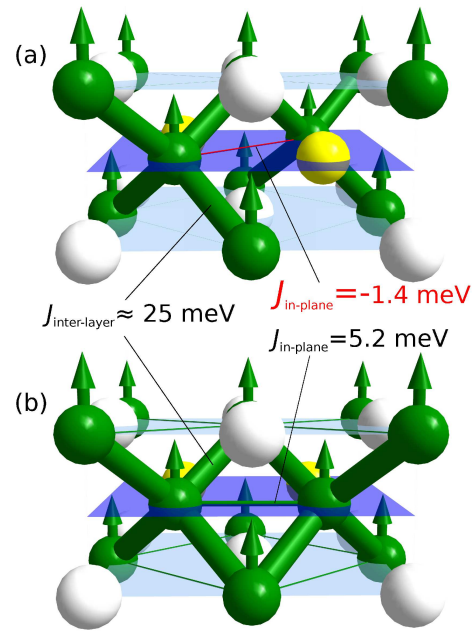}
\caption{(color online) Comparison of the  magnetic exchange
  coupling in the ``inverted'' FeGa/FeCu~(a) and in the
  most stable FeGa/[Fe$_{0.5}$Cu$_{0.5}$][Fe$_{0.5}$Cu$_{0.5}$]~(b)
  configurations. The atoms are arranged within Fe-Ga and Fe-Cu layers
  marked by light- and dark-blue horizontal planes, respectively. Fe, Cu and Ga atoms are shown
  as green, yellow and white spheres, respectively. Magnetic moments  
  are shown by arrows. The bond thickness reflects the strength of the exchange interaction. The inter-layer Fe-Fe
  interactions are dominating: ${J_{\rm inter\text-layer}\approx 25}$~meV (thick green bonds).  The
  in-plane interactions are negligibly small, except those in Fe-Cu planes.  
  Case (b) illustrates  the typical distinction from
  the ordered ``inverted'' structure: the random in-plane swap of one Fe and one
  Cu atoms which  brings  two Fe atoms closer to one
  another within the Cu-Fe plane. This alters the nearest in-plane Fe-Fe exchange from
  antiferromagnetic (${J_{\rm in\text-plane}=-1.4}$~meV, thin red bond) in case (a) to
  ferromagnetic ($J_{\rm in\text-plane}=5.2$~meV, thin green bond) in case (b).\label{FIG:jxc}}
\end{figure}

The subsequent calculations of the magnetic exchange coupling constants
$J_{ij}$ (Fig.~\ref{FIG:jxc}) of the classical Heisenberg model
(${H=-\sum_{i>j}J_{ij}\hat{\rm e}_i\hat{\rm e}_j}$, where $\hat{\rm e}_{i,j}$
are the unity vectors along the magnetization directions on local sites $i$ and $j$)
revealed the magnetic origin of both stabilization mechanisms responsible for the atomic 
rearrangement from the ``regular'' into the ``inverted'' phase and
for the chemical disorder within the Fe-Cu layers. 
Namely, the strong Fe-Fe inter-layer coupling 
(${J_{\rm inter\text-layer}\approx 25}$~meV between the adjacent Fe-Ga and Fe-Cu layers) 
keeps the whole system ferromagnetic.
This is in agreement with the high Curie temperature (798~K) measured in Ref.~\cite{GKK+12}.
Although the in-plane couplings appear to be an order of magnitude weaker, still, as we mentioned, 
their optimization plays a crucial role in the stabilization of the cubic phase. 
In the ordered ``inverted'' configuration (FeGa/FeCu) the nearest in-plane Fe atoms
tend to couple antiferromagnetically (${J_{\rm in\text-plane}=-1.4}$~meV).
This interaction works against the overwhelming ferromagnetic order already set by the
strong inter-layer coupling. 
Thus, the magnetic energy can be further reduced by
bringing the Fe atoms closer together as shown in Fig.~\ref{FIG:jxc}\,(b),
so that they couple ferromagnetically (${J_{\rm in\text-plane}=5.2}$~meV). 
Put in practical terms, this effect favors the formation of random Fe clusters 
within the Fe-Cu planes, which --in contrast to the ordered case-- 
can be more adequately described by the chemical disorder picture, i.\,e. by
implying the FeGa/[Fe$_{0.5}$Cu$_{0.5}$][Fe$_{0.5}$Cu$_{0.5}$] configuration. 

\begin{figure}
\centering
\includegraphics[clip,width=0.8\linewidth]{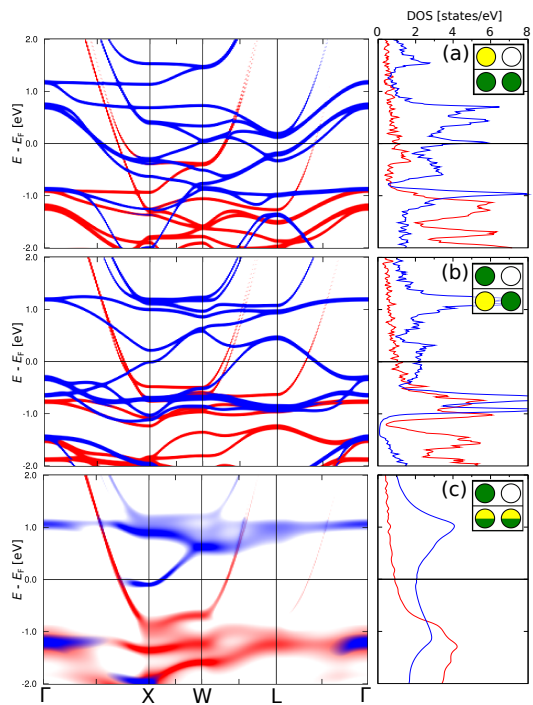}
\caption{(color online) Comparison of the  spin-resolved electronic band
  structures and related densities of states for the (a) ``regular'' cubic CuGa/FeFe, (b) ``inverse'' cubic FeGa/CuFe
  and (c) FeGa/[Fe$_{0.5}$Cu$_{0.5}$][Fe$_{0.5}$Cu$_{0.5}$]
  configurations. The majority- and minority-spin states are
  distinguished by red and blue, respectively.\label{FIG:bands}}
\end{figure}

The considerations presented above can be futher supported by comparing 
the electronic structures as shown on Fig.~\ref{FIG:bands}.
The instability of the electronic subsystem is typically related to the
strength of the DOS peaks in the vicinity of the Fermi energy.
In case of the ``regular'' CuGa/FeFe cubic system, a huge instability peak
at $E_{\rm F}$ (total DOS${(E_{\rm   F})\approx6.1}$~sts./eV) is produced by
the van-Hove singularity in the minority-spin channel at W-point of the Brillouin zone~(Fig.~\ref{FIG:bands}\,(a)).
By replacing Cu from the $4a$ (or $4b$) site will split the minority-spin states
at W-point far away from the Fermi energy, noticeably reducing the DOS peaks 
(see Fig.~\ref{FIG:bands}\,(b) and (c)). 
This step is related to the largest ($-170$~meV/f.u.) energy gain
attributed to the inter-layer magnetic exchange optimization discussed above.
Now by going from the ordered ``inverted'' Heusler variant (FeGa/CuFe) to the most
stable case of FeGa/[Fe$_{0.5}$Cu$_{0.5}$][Fe$_{0.5}$Cu$_{0.5}$], the DOS at $E_{\rm F}$ 
is further reduced (from 3.2 to 2.9~sts./eV).
Indeed, as we have shown above, the Fe-Cu disordered configuration benefit from
the in-plane magnetic optimization, and therefore it is by 40~meV/f.u. lower in 
energy compared to the ``inverted'' FeGa/CuFe system. 
We would like to point out, that very similar stabilization mechanisms characterized
by comparable energy scales can take place in other Fe$_2$-based cubic Heusler compounds. 
For example, for Fe$_2$CuAl and  Fe$_2$NiGe the
inter-plane exchange energy optimization (i.\,e. by going from the ``regular'' to the ``inverted''
cubic phase) gains  $-367$ and $-168$~meV/f.u., whereas the in-plane optimization
(due to Fe-Y in-plane disorder) contributes with $-27$ and
$-40$~meV/f.u., respectively to the energy.
Thus the stabilization mechanisms presented in this letter are rather general
within the group of Fe$_2$YZ materials.

To conclude, we emphasize that the presented analysis explains the stability
of the cubic phase in Fe$_2$YZ Heusler compounds, and provides
a clear explantion for the discrepancy between experimental results and theory.
The actual stabilizing mechanism appears to be the chemical disorder,
which optimizes the magnetic exchange coupling within
Fe-Y layers of the initially ordered FeZ/FeY cubic phase.
At the same time, the FeZ layers remain chemically ordered due to a large
difference (i.\,e.\@ atomic radius, valency, electronegativity etc.) 
between Fe and the main-group element Z. 
Thus, the most stable configurations can be written as 
FeZ/[Fe$_{0.5}$Y$_{0.5}$][Fe$_{0.5}$Y$_{0.5}$].
In general, the important prerogative enabling the chemical disorder is the ``inverted'' ordered
structure: as we have seen, the effect of the rearrangement from YZ/FeFe
into FeZ/YFe within the cubic phase is comparably efficient to 
the tetragonal distortion in YZ/FeFe. On the other hand, the first scenario allows
to further optimize the system by chemical disorder, whereas the
second one does not.

\acknowledgments
The authors thank to Sergey Medvedev (MPI Dresden) for the helpful discussions.
Financial support by the DFG project FOR 1464 “ASPIMATT” (1.2-A) is gratefully acknowledged.

\end{document}